\RequirePackage[l2tabu,orthodox]{nag}
\pdfoutput=1
\def\mytitle{Steering droplets on substrates with plane-wave wettability patterns and deformations}
\def\authors{Josua Grawitter and Holger Stark}

\documentclass[a4paper,oneside,10pt]{article}

\usepackage[utf8]{inputenc}
\usepackage[english]{babel}

\usepackage{charter}
\usepackage[T1]{fontenc}

\usepackage{amsmath}
\usepackage{amssymb}
\usepackage{mathtools}
\usepackage{comment}
\usepackage{textcomp}
\usepackage{pdfpages}
\usepackage{appendix}
\usepackage[style=base]{caption}
\usepackage{bm}
\usepackage{url}
\usepackage{pdflscape}
\usepackage[margin=15mm,tmargin=2cm,bmargin=2.5cm]{geometry}

\usepackage{setspace}

\usepackage{sectsty}
\sectionfont{\fontsize{12}{15}\selectfont}
\subsectionfont{\fontsize{10}{13}\selectfont}

\usepackage{csquotes}
\usepackage[numbers,sort&compress,comma]{natbib}

\usepackage{overpic}
\usepackage[normalem]{ulem}

\renewcommand{\vec}[1]{\bm{#1}}

\newcommand{\dif}{\,\mathrm{d}}
\newcommand{\latin}[1]{\textit{#1}}

\usepackage{colorprofiles}
\usepackage[a-2b,mathxmp]{pdfx}[2018/12/22]
\hypersetup{
    pdfstartview=,
    colorlinks=false,
    pdfborder={0 0 0},
    linktoc=all,
    plainpages=false,
    hypertexnames=false, 
}

\usepackage{cleveref}
\usepackage[rgb,hyperref]{xcolor}

\usepackage{lipsum}

\usepackage{microtype}

\title{\mytitle}
\author{\authors}
\begin{document}
\twocolumn[
{\LARGE \mytitle}
\vspace{20pt}
\newline{\Large \authors}\newline
\abstract
{
Motivated by strategies for targeted microfluidic transport of droplets, we investigate how sessile droplets can be steered toward a preferred direction using travelling waves in substrate wettability or deformations of the substrate.
To perform our numerical study, we implement the boundary-element method to solve the governing Stokes equations for the fluid flow field inside the
moving droplet.
In both cases we find two distinct modes of droplet motion.
For small wave speed the droplet surfes with a constant velocity on the wave, while beyond a critical wave speed a periodic wobbling motion occurs, the period of which diverges at the transition.
These observation can be rationalized by the nonuniform oscillator model and the transition described by a SNIPER bifurcation.
For the travelling waves in wettability the mean droplet velocity in the wobbling state decays with the inverse wave speed.
In contrast, for travelling-wave deformations of the substrate it is proportional to the wave speed at large speed values since the droplet always has to move up and down.
To rationalize this behavior, the nonuniform oscillator model has to be extended.
Since the critical wave speed of the bifurcation depends on the droplet radius, this dependence can be used to sort droplets by size.
}
\newline
\vspace{40pt}
]

\section{Introduction}

Liquid droplets on a solid substrate occur naturally during, \emph{e.g.}, rain and condensation \cite{alwazzan_condensation_2017,edalatpour_managing_2018}  or artificially in printing of ink \cite{varanakkottu_light_2016} or medical testing \cite{qi_mechanical_2019}.
Depending on their interaction with the substrate and gas phase, they either sit on a substrate, spread, contract, or even move laterally along the substrate~\cite{bonn_wetting_2009}.
Lateral motion can be induced by a variety of mechanisms;
through nonuniform electric fields (electrowetting)
\cite{teng_recent_2020}, gravity \cite{thiele_sliding_2002}, Marangoni advection \cite{baigl_photo_2012,venancio_digital_2014},
gradients in wettability \cite{chaudhury_how_1992,ichimura_light_2000,qi_mechanical_2019}, or deformations of the substrate itself \cite{vialetto_magnetic_2017,schiphorst_light_2018}.
Here, we focus on the latter two mechanisms.

In our previous work, we used the gradient in a wettability step to move a droplet forward using a feedback loop, which synchronizes
step and droplet motion~\cite{grawitter_steering_2021}.
However, due to surface imperfections at which droplets get pinned, such a motion can be difficult to realize in an experiment~\cite{bonn_wetting_2009}.
To overcome the pinning, a large wettability gradient is needed \cite{eral_contact_2013}, \emph{e.g.}, induced by structural changes of the substrate~\cite{lim_photoreversibly_2006}.

Deformations of the substrate itself also move a droplet\ \cite{karpitschka_liquid_2016}.
In Ref.~\cite{lv_substrate_2014} it has been demonstrated that capillary forces attract droplets towards regions with large inward curvature.
An example of this are pipettes.
It is possible to externally control such substrate deformations, \emph{e.g.}, using gels that reversibly swell and unswell in response to light with specific wavelengths~\cite{palagi_structured_2016,schiphorst_light_2018,rehor_photoresponsive_2021}.
Already more than three decades ago wetting of curved rigid substrates was investigated~\cite{gelfand_wetting_1987}.
In more recent work, elastic substrates were studied that deform in contact with a liquid droplet~\cite{duprat_elastocapillarity_2016,bico_elastocapillarity_2018,aland_unified_2021}.
Internal tension builds up as a substrate deforms, which changes its surface tension what is known as Shuttleworth effect~\cite{shuttleworth_surface_1950,andreotti_statics_2020}.
Here, we just consider curved \emph{rigid} substrates, where this effect does not occur.

\begin{figure}
\centering
 \includegraphics[width=\linewidth]{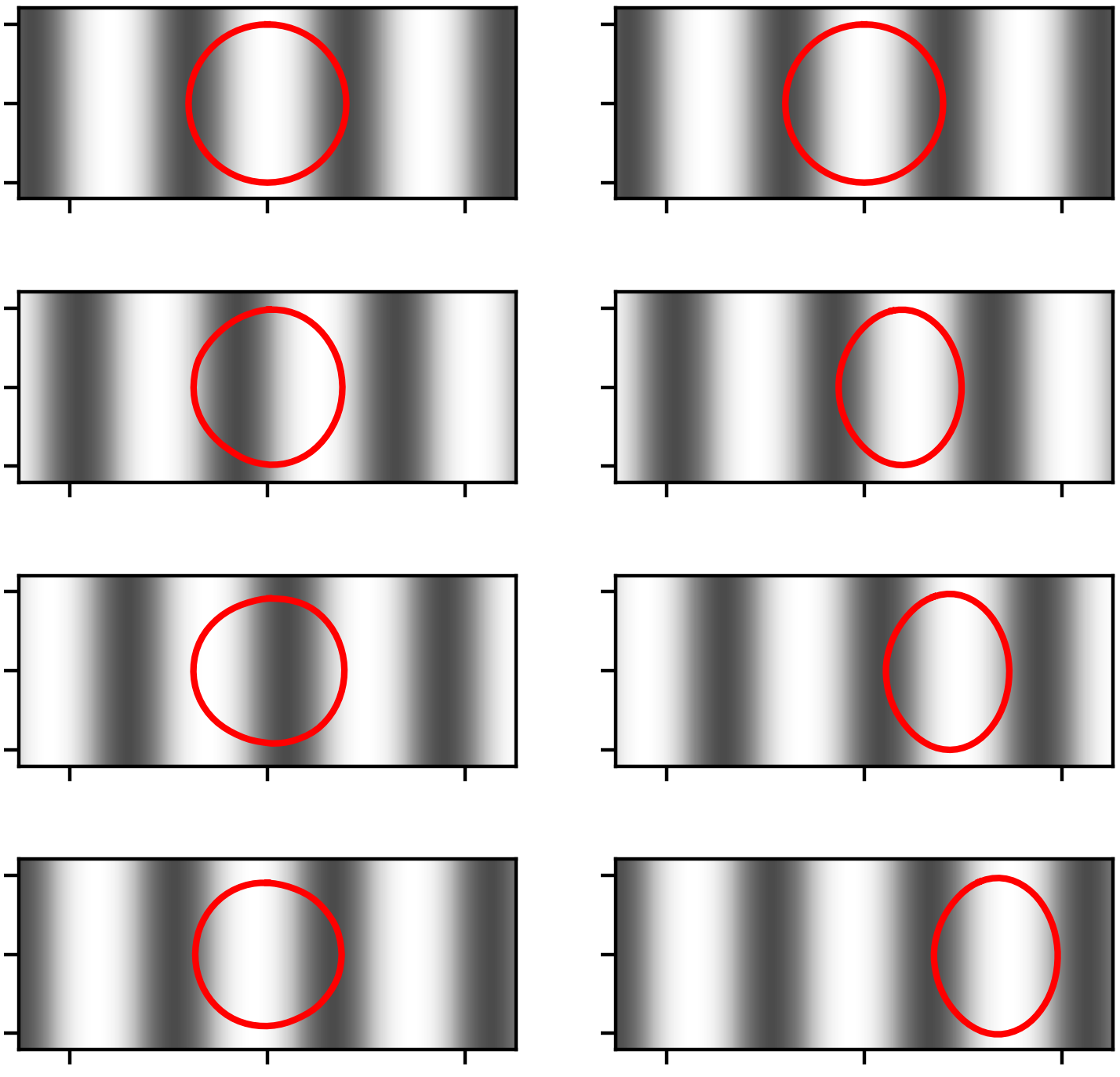}
 \caption{Time series (top to bottom) of snapshots for wettability waves travelling at different speeds: (left) \emph{wobbling} motion at large speeds and (right) \emph{surfing} motion of the droplet at small speeds.
The greyscale shading indicates wettability and the red lines indicate the contact line of the droplet.}
\label{fig.wettability_waves}
\end{figure}

In this article we compare two mechanisms for inducing droplet motion by imposing the spatio-temporal pattern of a travelling wave onto the substrate; first in wettability and thereafter in the height profile of the substrate.
This enables to steer the droplet into a prescribed direction.
Because the travelling wave is spatially periodic, the droplet is continuously driven out of equilibrium and cannot settle into a sessile state.
Instead, the droplet can be considered as a driven nonuniform oscillator \cite{strogatz_nonlinear_1994}, as we show, the driving force of which depends on its position relative to the travelling wave.
In Fig.~\ref{fig.wettability_waves} we display two series of snapshots that illustrate the two possible modes of motion, which we term wobbling (left in Fig.~\ref{fig.wettability_waves}) and surfing (right in Fig.~\ref{fig.wettability_waves}), respectively.
In the following we reformulate our formalism to apply the boundary element method (BEM)~\cite{pozrikidis_boundary_1992} to a wetting droplet, which we developed in Refs.~\cite{grawitter_steering_2021,grawitter_droplets_2021}, and adjust its boundary condition to account for either an imposed wettability pattern or a height profile of the substrate.
Because in our previous work we only studied a plane substrate~\cite{grawitter_steering_2021,grawitter_droplets_2021}, the method needed to be extended to curved substrates.

First, in Section \ref{sec_theory} we describe the theory of dynamic wetting for small droplets and our numerical approach based on the BEM.
Next, in Section \ref{sec_oscillator} we introduce the so-called \textit{nonuniform oscillator}, which we use as an analytic model to interpret the observed droplet dynamics.
We present and analyze the effect of travelling waves in wettability in Section~\ref{sec_wettability} and travelling-wave deformations in Section~\ref{sec_deformations}.
Finally, we comment on implications of our findings for the sorting of droplets by size in Section~\ref{sec_tuning} and conclude with Section~\ref{sec_conclusions}.

\section{Boundary element method for dynamic wetting}
\label{sec_theory}
We consider the motion of viscous droplets at small scales in the creeping flow regime, where viscous forces dominate, while inertia is negligible.
Accordingly, fluid flow within droplets are described by the Stokes equations and the incompressibility condition,
\begin{equation}
 \eta \nabla^2 \vec v=\nabla p
 ~\text{and}~
 \nabla\cdot \vec v = 0 \, .
 \label{eq_stokes}
\end{equation}
Here, we use fluid velocity field~$\vec{v}$, pressure~$p$, and shear viscosity~$\eta$.
In addition, appropriate boundary conditions have to be chosen, most importantly, describing forces acting at the fluid boundary.
Notably, the flow fields determined by the Stokes equations adapt instantaneously to the boundary conditions because the equations do not contain any time derivatives.
As a result, when describing the dynamics of the droplet boundary by a surface velocity field $\dot{\vec R}$, it is linked to a friction force field by the linear relation
\begin{equation}
  \vec K=-\vec G\dot{\vec R} \, ,
  \label{eq_ansatz}
\end{equation}
where $\vec G$ is a friction operator~\cite{doi_onsager_2011,zafferi_generic_2023}.
Since the Stokes equations are linear in the fluid velocity field $\vec v$,
 $\vec G$ is a linear operator that does not depend on $\vec v$.
After discretizing the droplet surface, $\vec G$ becomes a friction matrix and $\vec R$ and $\vec K$ are vectors.

Our goal is to determine $\vec K$, $\vec R$, and $\vec G$ explicitely for the mesh discretization of the droplet surface described below.
This will give us a systematic approach for calculating the motion of the droplet.
We start by introducing the discretized droplet surface, which specifies the vector $\vec R$, then formulate the boundary condition at the interfaces to determine the generalized friction force~$\vec K$, and finally derive the friction matrix~$\vec G$.
In addition, we nondimensionalize our quantities in Section~\ref{sec_nondimensionalization}.

\subsection{Mesh discretization}

As noted above, our aim is to reduce the droplet motion to the dynamics of its boundary consisting of the liquid-substrate and liquid-gas interfaces as well as the three-phase contact line, which requires special consideration.
To treat the droplet dynamics numerically, we describe the entire droplet surface by a triangular mesh.
Each triangle consists of three vertices and three edges, which it shares with its direct neighbour triangles.
As long as the triangular mesh stays intact, the shape of the droplet is completely specified by the positions of the vertices~$\vec r_i$, which define the configuration vector~$\vec R=(\vec r_1, \vec r_2, \vec r_3,\ldots)$.

In general, the part of the surface force field acting on the droplet at the liquid-solid and liquid-gas interactions derives from the functional derivative of a surface energy w.r.t.~the configuration field $\vec R$.
However, since the droplet surface is discretized, the functional derivative becomes gradients w.r.t.~the vertex positions.
To continue, we construct an expression for the surface energy using surface integrals and discretize the integrals for our mesh in the following part.

\subsection{Droplet free energy and discretization}
The free energy of an incompressible liquid droplet at constant temperature can be written as
\begin{equation}
\mathcal{F}_0 = \oint_{A} \gamma(\vec r) \dif^2\vec r + \mathrm{const.} \, ,
\end{equation}
where $\gamma$ is the surface tension and the area~$A$ encompasses all the surfaces of the droplet.
The bulk free energy is kept constant here.
The surface integral can be split up into two parts: The liquid-gas (lg) interface and the solid-liquid interface (sl) with the respective surface tensions $\gamma_\mathrm{lg}$ and $ \gamma_\mathrm{sl}$.
The solid-gas interface of the substrate not in contact with the droplet contributes to the free energy with surface tension $\gamma_\mathrm{sg}$.
Whenever part of this interface is covered by the droplet, the surface energy changes by $\gamma_\mathrm{sl} -  \gamma_\mathrm{sg}$.
Thus we obtain
\begin{equation}
\mathcal{F}_0 = \int_{A_\mathrm{lg}} \gamma_\mathrm{lg} \dif^2\vec r + \int_{A_\mathrm{sl}}  (\gamma_\mathrm{sl} -  \gamma_\mathrm{sg}) \dif^2\vec r + \mathrm{const.} \, ,
\end{equation}
where $A_\mathrm{lg}$ and $A_\mathrm{lg}$ are the areas of the respective interfaces.
Young's law for the equilibrium contact angle~$\theta_\mathrm{eq}$ is a force balance in the substrate plane that states that \(\gamma_\mathrm{sl}-\gamma_\mathrm{sg}=-\gamma_\mathrm{lg}\cos\theta_\mathrm{eq}\).%
\footnote{In fact, Young's law follows from minimizing \(\mathcal{F}_0\) under a volume constraint.}
Using it together with $\gamma_\mathrm{lg}=\mathrm{const.}$, we obtain
\begin{equation}
\mathcal{F}_0 = \gamma_\mathrm{lg} A_\mathrm{lg} - \gamma_\mathrm{lg} \int\limits_{A_\mathrm{sl}} \cos\theta_\mathrm{eq}(\vec r) \dif^2\vec r + \mathrm{const.} \, ,
\end{equation}
where $\theta_\mathrm{eq}$ is determined by the \emph{local} value of $\gamma_\mathrm{sg}-\gamma_\mathrm{sl}$.
Because we want to study movable droplets on substrates with nonuniform wettability, the remaining integral cannot be further simplified here.

We now turn our attention toward the constraints on the shape of the droplet.
They are two-fold:
First, the volume of the droplet is conserved and, second, the normal forces on the liquid-solid interface sum up to zero since any fluid forces are compensated by the forces exerted by the rigid substrate that resists any elastic deformation.
Using the method of Langrangian multipliers, each constraint gives rise to an additional term in the energy functional:
\begin{equation}
\mathcal{F} = \mathcal{F}_0 + p_0(V_0-V)+
\int\limits_{A_\mathrm{sl}} \lambda(\vec x)
\left[z - h(\vec x)\right]
\dif^2\vec r
\label{eq.F}
\end{equation}
Here, the second term on the right-hand side constrains the droplet volume to the value $V_0$ and the pressure $p_0$
serves as the Langrange multiplier.
The third term firmly connects the vertical position $z$ of the liquid-solid interface to the
height variations $h(\vec x)$ of the substrate written in Monge representation with $\vec x = (x,y)$.
In the case of a flat substrate, $h=0$, this term reduces to
\begin{equation}
 \int\limits_{A_\mathrm{sl}} \lambda(x,y) z \dif x\dif y \, ,
 \label{eq_rigid_constraint_flat}
\end{equation}
which guarantees that the solid-liquid interface~$A_\mathrm{sl}$ is defined by $z=0$.
Finally, note that the volume of the droplet can be expressed by the surface integral
\begin{equation}
 V=\frac{1}{3}\oint \vec r\cdot \vec n\dif^2\vec r~ \, ,
\end{equation}
as one verifies immediately with Gauss's theorem.

To discretize all the surface integrals of the free energy $\mathcal F$, we use the triangular mesh of the droplet surface and sum up the contributions from each triangle so that
\begin{equation}
 \int_A f(\vec r)\dif^2 \vec r \approx \sum_i \int_{A_i} f(\vec r) \dif^2 \vec r~.
\end{equation}
Here, we have to distinguish between functions~$f$, the values of which are known for any position $\vec r$ and those functions, the values of which are only known at the vertices.
In the first case, an example is  $\cos\theta_\mathrm{eq}$, we use Gaussian quadrature with 4~sample values of the prescribed profile to calculate the integral on each triangle.
The sample values are taken at positions (1/3, 1/3, 1/3),  (3/5, 1/5, 1/5), (1/5, 3/5, 1/5), and (1/5, 1/5, 3/5) in barycentric coordinates of the triangle and weighted by --27/48, 25/48, 25/48, and 25/48, respectively~\cite{katsikadelis_gauss_2016}.
In the second case, an example is the calculation of the volume $V$, we linearly interpolate between the values at the vertices.
If $f$ has a constant value $f_i$ on the area of the triangle, for example, in the case of $\gamma_\mathrm{lg}$, we simply use $A_i f_i$.

\subsection{Method of Lagrange multipliers}
\label{sec_lagrange}
After having discretized all contributions to $\mathcal F$, we obtain terms as a function of the configuration vector~$\vec R$, including all the constraint terms, which we write as $\lambda_j g_j(\vec R)=0$.
For convenience, we label $\lambda_0=p_0$ and assign the remaining multipliers to the rigidity constraint.
Thus, Eq.~(\ref{eq.F}) becomes
\begin{equation}
\mathcal{F} = \mathcal{F}_0 + \sum\limits_{i=0}^N \lambda_i g_i(\vec R) + \mathrm{const.}
\label{eq.F2}
\end{equation}

We discretize the surface integral of the rigidity constraint in Eq.~(\ref{eq.F}),
\begin{equation}
\sum_{i=1}^{N} \lambda_i g_i(\vec R) =
\sum_{i=1}^{N} \lambda_i [z_i - h(x_i, y_i)] \, ,
\end{equation}
where we subsume the area~$A_i$ surrounding each vertex into the $\lambda_i$, respectively, and we sum over all~$N$ vertices on the solid-liquid interface.
The force due to this constraint is guaranteed to be normal to the substrate, since $\nabla_i g_i=\vec n_i$, as shown in Appendix~\ref{app_normal}, and therefore equivalent to the desired rigidity constraint.%
\footnote{In the case of a flat substrate placed at a height~$h=0$, we have $\vec n_i=\vec e_z$ and so the constraint reduces to $\lambda_i g_i=\lambda_i z_i$, which is aquivalent to Eq.~(\ref{eq_rigid_constraint_flat}).}
This construction greatly simplifies the procedure below because $\nabla_{\vec R} g_i$ is easily accessible.

The purpose of the Lagrange multipliers~$\lambda_i$ is to constrain the force $-\nabla_{\vec R} \mathcal{F}$ such that it does not have a component perpendicular to the manifold defined by the constraints~$g_i(\vec R)=0$, otherwise the constraints would be violated.%
\footnote{This is commonly referred to as the \emph{principle of virtual work}.}
Therefore, the condition
\begin{equation}
  (\nabla_{\vec R} g_i) \cdot (\nabla_{\vec R} \mathcal{F}) = 0
\end{equation}
must hold.
Using Eq.~(\ref{eq.F2}) for $\mathcal{F}$ in this condition with
\begin{equation}
\nabla_{\vec R} \mathcal{F} = \nabla_{\vec R} \mathcal{F}_0 + \sum_i \lambda_i \nabla_{\vec R} g_i \, ,
\label{eq.nabla_F}
\end{equation}
we obtain a set of linear equations for the Lagrange multipliers $\lambda_i$ (including $\lambda_0$),
\begin{equation}
 -\sum_j (\nabla_{\vec R} g_i) \cdot (\nabla_{\vec R} g_j) \lambda_j =
 (\nabla_{\vec R} g_i) \cdot (\nabla_{\vec R} \mathcal{F}_0) \, ,
\end{equation}
which we solve numerically.
With this, $\mathcal F$ is known up to a constant.%
\footnote{%
Note that we can reuse the gradients $\nabla_{\vec R} g_i$ from above in calculating $\nabla_{\vec R} \mathcal{F}$ in Eq.~(\ref{eq.nabla_F}).
Note also that in our previous study Ref.~\cite{grawitter_steering_2021} when we studied a droplet on a rigid flat substrate there was no need to calculate the constraint forces due to the rigid substrate because in that case we calculated the force on each vertex individually and imposed the volume constraint only on the resulting velocity field.
That approach proved inapplicable for our current study.
}

\subsection{Boundary condition and dynamic equation}
\label{sec_boundary}
The boundary condition of the Stokes equations, Eq.~(\ref{eq_stokes}), is the balance of forces acting at each point on the interfaces of the droplet.
There are three contributions to this force balance.
First, the traction of the liquid on the interface, $\vec{\sigma}\vec n$, where $\sigma_{ij}=-p\delta_{ij}+\eta(\partial_i v_j + \partial_j v_i)$ is the stress tensor, $\vec n$ the unit normal vector, and $\delta_{ij}$ is the Kronecker symbol.
For the following argumentation, we split the stress tensor into a part $-p_0\vec{n}$, where $p_0$ is the spatially uniform static pressure due to the volume constraint, and a spatially varying part~$\tilde{\vec \sigma}\vec n= \vec \sigma \vec n + p_0\vec n$.
Second, the traction of the substrate at the substrate-liquid interface.
It consists of the constraint force $\vec f_\mathrm{rigid}$, with which the substrate resists deformations induced by the droplet, and $\vec f_\mathrm{deform}$, which acts on the liquid when the substrate deforms according to the prescribed height profile $h$.
And third, the traction due to the surface tension of both interfaces which includes the Young force acting on the contact line and which we denote here by $-\delta \mathcal F_0/\delta \vec r$, \latin{i.e.}, the functional derivative of the surface free energy w.r.t.~the surface parameterization.
Note that we neglect the vapour pressure of the gas phase surrounding the droplet, here and in the following.
In total, we have the force balance at each point of the droplet surface,
\begin{equation}
\vec{\tilde \sigma} \vec n - p_0\vec n  = - \frac{\delta \mathcal F_0}{\delta \vec r} + \vec f_\mathrm{rigid} + \vec f_\mathrm{deform}  \, ,
\label{eq.balance}
\end{equation}
where $\vec f_\mathrm{rigid} + \vec f_\mathrm{deform}$ is zero at the gas-liquid interface.
Note that $\vec f_\mathrm{deform}$ is due to the boundary conditions at the liquid-substrate interface, which include continuity of the normal velocity component and a slip boundary condition of the tangential component, as we will specify in Section~\ref{subsub.ls_friction}.

The first two terms on the r.h.s.~of Eq.~(\ref{eq.balance}) together with $p_0 \vec n$ can be rewritten as $-\delta \mathcal F/\delta \vec r$ using the Lagrangian constraints introduced in Eq.~(\ref{eq.F}).
So we obtain
\begin{equation}
 \vec {\tilde\sigma} \vec n = - \frac{\delta \mathcal F}{\delta \vec r} + \vec f_\mathrm{deform}  \, .
\label{eq.balance2}
\end{equation}
To illustrate this force balance, we consider the equilibrium case of the plane surface, $\vec f_\mathrm{deform} = \vec 0$, and no
spatially varying stress, $\vec {\tilde\sigma} \vec n= \vec0$, such that $\delta \mathcal F/\delta \vec r = \vec 0$.
On the liquid-gas interface, $\delta \mathcal F_0/\delta \vec r = -2\kappa\gamma_\mathrm{lg}\vec n$~\cite{deckelnick_dziuk_elliott_2005}
so that $p_0$ becomes the Laplace pressure
\begin{equation}
p_L = -2\kappa\gamma_\mathrm{lg}\, ,
\end{equation}
where $\kappa$ is the mean curvature of the droplet, which is here negative.

We now rephrase boundary condition~(\ref{eq.balance2}) for the discretized droplet surface.
At each vertex $i$ the traction~$\tilde{\vec \sigma}_i\vec n_i$ from the fluid is multiplied by area~$A_i$ and collected in the friction vector
\begin{equation}
\vec K = (-\vec{\tilde \sigma n}_1 A_1, -\vec{\tilde \sigma n}_2 A_2, -\vec{ \tilde\sigma n}_3 A_3, \ldots) \, ,
\label{eq_K_definition}
\end{equation}
Similarly, we collect the substrate traction $\vec f_\mathrm{deform}$ in a force vector
\begin{equation}
 \vec F = (\vec f_\mathrm{deform}^{(1)} A_1, \vec f_\mathrm{deform}^{(2)} A_2, \vec f_\mathrm{deform}^{(3)} A_3, \ldots) \, .
\label{eq_F_definition}
\end{equation}%
Our method to calculate $\vec F$ is described in detail in Section~\ref{sec_friction}.
As usual, the gradient of $\mathcal F$ describes a force $-\nabla_{\vec R} \mathcal{F}$,
which takes the place of the functional derivative $-\delta \mathcal F/\delta \vec r$.
Thus, in analogy to condition~(\ref{eq.balance2}), we have
\begin{equation}
   \vec K+\vec F -\nabla_{\vec R} \mathcal{F}=0~.
\label{eq_force_balance_first}
\end{equation}
Finally, with the ansatz $\vec K=-\vec G\dot{\vec R}$ from Eq.~(\ref{eq_ansatz}), which we will motivate further in the following section, the relaxational dynamics of the surface is given by
\begin{equation}
\vec {\dot{R}} = -\vec{G^{-1}}(\nabla_{\vec R} \mathcal{F}-\vec F)~.
\label{eq_force_balance}
\end{equation}
Here, we observe that for a stationary substrate with $\vec F=\vec 0$, a local minimum of $\mathcal F$ corresponds to a stationary droplet since then $\vec {\dot{R}} = \vec 0$.

\subsection{Generalized friction}
\label{sec_friction}
Now, we specify the friction matrix~$\vec G$ such that all sources of friction related to the droplet are accounted for.
Three types of friction are relevant in our setup.
First, there is shear friction or viscous shear stresses in the droplet fluid, when neighboring fluid layers move relative to each other.
Second, there is slip friction between the substrate and the directly adjacent fluid layer.
And finally, there is friction of the moving contact line.
The latter should naturally arise from the first and second mechanisms.
However, so far this has proved difficult to show analytically~\cite{bonn_wetting_2009} and also to implement numerically due to the flow singularity that arises at the contact line~\cite{huh_hydrodynamic_1971}.
Therefore, we use a hydrodynamically motivated friction law for the contact line.
It is closely related to the Cox-Voinov law~\cite{voinov_hydrodynamics_1976}, which has successfully been matched to experiments, \latin{e.g.}, in Ref.~\cite{deruijter_contact_1997}.

\subsubsection{Shear friction}
We start with shear friction in the droplet fluid.
The solution of the Stokes equations within a given domain is equivalent to a boundary-integral equation that relates
flow velocity $\vec v(\vec r)$ to surface forces $\vec{ \sigma}({\vec r}) \vec n$ and force/source dipoles at the domain surface \cite{pozrikidis_boundary_1992},
 \begin{multline}
 \frac{\alpha}{4\pi}\vec v(\vec r_0)=\oint \vec O(
 \vec{r}_0-\vec{r}
 )
  \vec{ \sigma(\vec r) n} \dif^2\vec r
 \\-\oint \vec v(\vec r)\cdot \vec T(
 \vec{r}_0-\vec{r}
 ) \vec n \dif^2\vec r \, ,
 \label{eq_boundary_integrals}
\end{multline}
where $\alpha$ is the inward \emph{solid} angle, $ \vec O$ the Oseen tensor, and $\vec T$ the associated stress tensor.
For the surface velocities relevant here,  $\alpha=2\pi$ at the fluid interface and $\alpha=2\theta$ at the contact line, while inside the droplet $\alpha=4\pi$.
We note that $\vec{\sigma n}$ in Eq.~(\ref{eq_boundary_integrals}) can be replaced by $\vec{\tilde \sigma n}$ since a constant pressure does not contribute to the integral as shown in Appendix~\ref{app_pressure}.
To discretize the boundary-integral equation, we proceed as in our previous work \cite{grawitter_steering_2021}.
We assign to each surface vertex parts of the surrounding triangles such that the resulting polygonal cells with area $A_j$ do not overlap and cover the whole region of integration.
The result is a set of linear equations for the vertex velocities $\vec v_j$,
\begin{equation}
\sum_j c_i \delta_{ij}  \vec{v}_j = \sum_j X_{ij} \vec{\tilde \sigma n}_j - \sum_j Y_{ij} \vec{v}_j~.
\label{eq.bem}
\end{equation}
Recalling the definition of $\vec K$ in Eq.~(\ref{eq_K_definition}) and introducing block matrices, which we specify in Appendix~\ref{app_block_matrices},
\begin{equation}
 \vec X = \begin{pmatrix}
           X_{11}/A_1 & X_{12}/A_2 & \ldots\\
           X_{21}/A_1 & X_{22}/A_2 & \ldots\\
           \ldots & \ldots & \ldots\\
          \end{pmatrix},
\end{equation}
\begin{equation}
 \vec Y = \begin{pmatrix}
           Y_{11} & Y_{12} & \ldots\\
           Y_{21} & Y_{22} & \ldots\\
           \ldots & \ldots & \ldots\\
          \end{pmatrix}
\enspace \mathrm{and} \enspace
 \vec C = \begin{pmatrix}
           c_1 & 0 & \ldots\\
           0 & c_2 & \ldots\\
           \ldots & \ldots & \ldots\\
          \end{pmatrix} \, ,
\end{equation}
we are able to rewrite Eq.~(\ref{eq.bem}) as
\begin{equation}
\vec C \vec {\dot R} =
    -\vec X \vec K
    -\vec Y \vec {\dot R}~.
\end{equation}
or as
\begin{equation}
\vec K = -\vec X^{-1}(\vec C + \vec Y)\vec {\dot R} \, ,
\label{eq.friction_shear}
\end{equation}
where $\vec X^{-1}(\vec C + \vec Y)$ is the shear contribution to the friction matrix.

\subsubsection{Liquid-substrate friction}
\label{subsub.ls_friction}
We now turn to friction between liquid and substrate due to a non-zero slip velocity.
The conventional slip boundary condition is $l_\mathrm{s} \vec{P}_t \vec \sigma\vec n=\eta \vec{P}_t \vec v$, where $l_\mathrm{s}$ is the slip length and $\vec{P}_t = \vec{1} - \vec{n} \otimes \vec{n}$ the projection operator on the tangential plane.
Note that the surface normal $\vec n$ points out of the fluid according to Ref.~\cite{bolanos_derivation_2017} and that $\vec P_t \vec{\sigma n}=\vec P_t \tilde{\vec{\sigma}}\vec n$.
However, since we are interested in moving substrates, the fluid motion relativ to the local substrate velocity $\vec v_\mathrm{subs}$ and the deforming traction~$\vec f_\mathrm{deform}$ become relevant and we have $l_\mathrm{s}\vec{P}_t (\tilde{\vec \sigma}\vec n - \vec f_\mathrm{deform}) = \eta \vec{P}_t (\vec v - \vec v_\mathrm{subs})$.
On the discretized solid/fluid interface of the substrate using Eqs.~(\ref{eq_K_definition}) and (\ref{eq_F_definition}), the boundary condition becomes
\begin{equation}
 \vec P_\mathrm{||}(\vec F + \vec K)
   =-\vec Q \vec P_\mathrm{||}(\vec {\dot R} - \vec V_\mathrm{subs}) \, .
\label{eq.friction_subs}
\end{equation}
Here, $\vec P_\mathrm{||}: \mathbb R^{3n} \to \mathbb R^{2m}$ is the projection operator from all the $n$ droplet vertices onto the tangential space of the substrate with $m$ vertices.
For consistency, we defined $\vec V_\mathrm{subs}=(\vec v_\mathrm{subs}^{(1)},\vec v_\mathrm{subs}^{(2)},\ldots)$ as the vector of the substrate velocities at the vertices, which also includes the vertices on the liquid-gas interface with $\vec v_\mathrm{subs}^{(i)}=\vec 0$.
Finally,
\begin{equation}
Q_{ij}=\frac{\eta}{l_\mathrm{s}} A_i \delta_{ij}
\end{equation}
is the slip contribution to the friction matrix, where $\delta_{ij}$ is the Kronecker symbol and $i, j$ refer to vertices on the substrate.
For the parallel components of vector $\vec F$ in Eq.~(\ref{eq_force_balance}), we can now write $\vec P_\mathrm{||}\vec F = \vec Q \vec P_\mathrm{||} \vec V_\mathrm{subs}$ so that $\vec K$ remains purely linear in $\vec {\dot R}$ as we have stated in Eq.~(\ref{eq_ansatz}).

So far, we have not addressed the continuity condition for the normal velocity component, $\vec{v} \cdot \vec{n} = \vec v_\mathrm{subs} \cdot \vec n$, at the liquid-substrate interface, which we already mentioned in Section~\ref{sec_boundary}.
Similar to $\vec P_\mathrm{||}$, it is convenient to introduce the projection operator onto the normal space of the substrate vertices $\vec P_{\bot} :\mathbb R^{3n}\to \mathbb R^{m}$.
The boundary condition can then be restated for our discretized mesh as $ \vec P_{\bot} \vec {\dot R}=\vec P_{\bot}\vec V_\mathrm{subs}$.
It directly determines the normal components of the vertex velocities at the substrate, which are contained in  $\vec {\dot R}$.
However, the condition also means that the substrate pushes or pulls on the liquid as it deforms its height profile~$h$ and the liquid resists with friction.
We have already introduced the traction forces due to the substrate deformations.
Their normal components are contained in the vector $\vec F_\bot=\vec P_\bot^T \vec P_\bot \vec F$, which has to be calculated numerically at the same time as those components of $\vec {\dot R}$ which are still unknown.

\subsubsection{Contact-line friction}
\label{subsub.contact_friction}
Finally, we address the friction of the contact line.
In 1964 Moffatt solved the Stokes flow at the contact line of a moving wedge filled with a viscous liquid and found an analytic expression for the traction~$\vec{\sigma n}$ along the liquid-gas interface~\cite{moffatt_viscous_1964},
\begin{equation}
\vec{n\cdot \sigma n} =
\frac{2\eta v_\mathrm{cl}}{H} \frac{\sin^2 \theta}{\theta - \sin \theta \cos \theta} \, .
\label{eq_moffatt}
\end{equation}
Here, $H$ is the distance of the interface from the substrate, $\theta$ the contact angle, and $v_\mathrm{cl}$ the tip velocity of the wedge along the substrate.
Again, if the substrate moves, we have $v_\mathrm{cl}=v-v_\mathrm{subs}$ 
and its deforming traction~$\vec f_\mathrm{deform}$, which modify Eq.~(\ref{eq_moffatt}).
The velocity lies in the tangential plane of the substrate and is normal to the contact line.
To proceed, we integrate the traction
over height~$H$ along the liquid/gas interface from a microscopic length $l_\mathrm{s}$, \latin{i.e.}, the slip length, up to a mesoscopic length $\zeta$
\begin{equation}
\int_{l_\mathrm{s}}^\zeta \vec{n\cdot \sigma n} \dif H = 2\eta \ln\left(\frac{\zeta}{l_\mathrm{s}}\right) \frac{\sin^2 \theta}{\theta - \sin \theta \cos \theta}v_\mathrm{cl}~.
\end{equation}
In the discretized representation of the droplet surface, we choose for the integral the surface force $\vec \sigma_i \vec n_i$ assigned to the vertex $i$ on the contact line, which is contained in the friction vector $\vec K$, \latin{i.e.},
\begin{equation}
\int_{l_\mathrm{s}}^\zeta \vec{n\cdot \sigma n} \dif H
=
\zeta \vec{n_i\cdot\sigma n}_i
~.
\label{eq_moffatt_integral}
\end{equation}
For the contact line, we therefore have the additional contribution
\begin{equation}
 \vec P_\mathrm{cl} (\vec F + \vec K)
 =-\vec M  \vec P_\mathrm{cl}  ( \vec{\dot R}-\vec V_\mathrm{subs}) \, ,
\label{eq.friction_contact}
\end{equation}
where $\vec P_\mathrm{cl}:\mathbb R^{3n}\to\mathbb R^{k}$ is the projection operator on the direction normal to the contact line
and tangential to the substrate, where $k$ is the number of vertices on the contact line.
Obviously, these vertices contribute with
\begin{equation}
M_{ij}= 2\eta \frac{A_i}{\zeta}\ln\left(\frac{\zeta}{l_\mathrm{s}}\right) \frac{\sin^2 \theta_i}{\theta_i - \sin \theta_i \cos \theta_i} \delta_{ij}
\label{eq_cl_friction_matrix}
\end{equation}
to the friction matrix.
Here we use the Kronecker symbol~$\delta_{ij}$, which means that $\vec M$ is diagonal.
As in Section~\ref{subsub.ls_friction}, we identify $\vec P_\mathrm{cl} \vec F=\vec M  \vec P_\mathrm{cl} \vec V_\mathrm{subs}$ as contribution to $\vec F$ in Eq.~(\ref{eq_force_balance}).
Therefore, in total we find
\begin{equation}
\vec F = (\vec P_\mathrm{||}^T \vec Q \vec P_{||} +\vec P_\mathrm{cl}^T \vec M \vec P_\mathrm{cl}) \vec V_\mathrm{subs} + \vec F_\bot
\label{eq.Fspecify}
\end{equation}
where $T$ indicates matrix transposition.

Finally, we are able to formulate the full set of equations.
Collecting the contributions from Eqs.~(\ref{eq.friction_shear}), (\ref{eq.friction_subs}), and (\ref{eq.friction_contact}) linear in $\vec{\dot R}$, the total friction matrix~$\vec G$ is thus given by
\begin{equation}
 \vec G =
 \vec X^{-1}(\vec C + \vec Y)
 + \vec P_{||}^T \vec Q \vec P_{||}
 +  \vec P_\mathrm{cl}^T \vec M \vec P_\mathrm{cl}~.
\end{equation}
By inverting $\vec G$, we arrive at the dynamical equation for $\vec R$,
\begin{equation}
 \vec{\dot R} = -\vec G^{-1}\vec K \, ,
 \label{eq_ansatz2}
\end{equation}
with $\vec K$ given by the force balance, Eq.~(\ref{eq_force_balance_first}), which we further specify here as
\begin{equation}
 \vec K = \nabla_{\vec R}\mathcal F - (\vec P_\mathrm{||}^T \vec Q \vec P_{||} +\vec P_\mathrm{cl}^T \vec M \vec P_\mathrm{cl}) \vec V_\mathrm{subs} - \vec F_\bot
 ~.
 \label{eq_force_balance_complete}
\end{equation}
using Eq.~(\ref{eq.Fspecify}).
We solve Eq.~(\ref{eq_ansatz2}) numerically to find the unknown components of $\vec {\dot R}$ and $\vec F_\bot$.
They include all components of $\vec {\dot R}$ corresponding to vertices on the liquid-gas interface and to the tangential motion of the substrate vertices, as well as all non-zero components of $\vec F_\bot$.
They are connected to the normal motion of the substrate vertices.

\subsection{Nondimensionalization and parameters}
\label{sec_nondimensionalization}
We choose characteristic units, which derive from the motion of the droplet surfaces.
First, as a unit of length~$l$, we choose the initial radius~$R_0$ of the liquid-solid interface of the droplet.
The dominant time scale for the deformation of the droplet is determined by the motion of the contact line.
A good estimate for the speed of the contact line was derived by Voinov in Ref.~\cite{voinov_hydrodynamics_1976} using Eqs.~(\ref{eq_moffatt}) and (\ref{eq.balance}).
He ultimately arrives the Cox-Voinov law
\begin{equation}
 v_\mathrm{cl}=\frac{\gamma_\mathrm{lg}}{9\eta\ln(\zeta/l_\mathrm{s})}(\theta^3-\theta_\mathrm{eq}^3) \, .
\end{equation}
From the prefactor of the Cox-Voinov law and $R_0$, we obtain an inherent time scale
\begin{equation}
\tau=9\gamma_\mathrm{lg}^{-1}\eta R_0\ln(\zeta/l_\mathrm{s}) 
\end{equation}
for the motion of the contact line.
Furthermore, a characteristic force follows from combining the liquid-gas surface tension~$\gamma_\mathrm{lg}$ and $R_0$ as $f_c=\gamma_\mathrm{lg} R_0$.
Now, writing Eq.~(\ref{eq_stokes}) in nondimensional form using these characteristic units, gives rise to a dimensionless viscosity $\tilde \eta = [9\ln(\zeta/l_\mathrm{s})]^{-1}$.
Thus, our results become independent from specific values of $\gamma_\mathrm{lg}$ and $R_0$.
They only depend on the ratios $\tilde \eta$, $\zeta /R_0$ and $l_\mathrm{s} / R_0$, for which we choose appropriate values.

As in Ref.~\cite{grawitter_steering_2021} the parameters of our system are matched to droplets from a 90\%-glycerol/10\%-water mixture
for which de Ruijter\ \emph{et\ al.}~\cite{deruijter_contact_1997} measured in experiments $\ln(\zeta/l_\mathrm{s})=44$,
$\eta=209\,\mathrm{mPa}\,\mathrm{s}$, $\gamma_\mathrm{lg}=65.3\,\mathrm{mN}\,\mathrm{m}^{-1}$, $l_\mathrm{s}=1\,\mathrm{nm}$,
and mass density $\rho=1.24\,\mathrm{g}\,\mathrm{ml}^{-1}$.
To determine a value for $\zeta$ needed in the friction matrix of Eq.~(\ref{eq_cl_friction_matrix}), we fitted the experimental data for the
droplet relaxation in Ref.~\cite{deruijter_contact_1997} in our simulations and obtained $\zeta=0.17\,R_0$.
All simulations are performed using a surface mesh of 1199~vertices which, in their initial configuration, have an average distance of
$0.094\,R_0$ from the nearest neighbor.
In the following, we use  $R_0=100\,\mu\mathrm{m}$.

From the characteristic parameters, we determine the Reynolds number as
\begin{equation}
\mathrm{Re}=\frac{\rho R_0^2 \tau^{-1}}{\eta}=\frac{\rho R_0 \gamma_\mathrm{lg}}{9\eta^2\ln(\zeta/l_\mathrm{s})}
= 5\cdot10^{-4} \, ,
\end{equation}
which matches the assumption of negligible inertia inherent in Eq.~(\ref{eq_stokes}).

\section{Driven droplets as nonuniform oscillators}
\label{sec_oscillator}
Our droplets are driven by travelling waves either in wettability or in deformations of the substrate.
For small wave speeds they are strongly connected to and surf on the wave, while for large speeds they can no longer follow and perform a wobbling motion.
Before we describe the phenomenology of both substrate types, we introduce and study a simplified model for the motion of a droplet under a spatially periodic external stimulus.
Neglecting inertia, we start with a simplified force balance for the center of mass~$y$ of the droplet, $-\Gamma \dot y + F(y,t) = 0$, where $\Gamma$ is a friction constant and $F$ is the effective force exerted by the substrate, on which the droplet is moving.
The travelling wave pattern is modeled by the effective force
\begin{equation}
F(y,t)=-F_0\sin\left(\frac{2\pi (y - v_\mathrm{wave} t)}{\lambda}\right)
\end{equation}
with wavelength $\lambda$ and wave speed $v_\mathrm{wave}$.
It models that there are specific positions in the travelling wave, $y - v_\mathrm{wave} t = n \pi$ with $n=0,1,\ldots$, where
no force is excerted on the droplet.
For our specific wave patterns, these are the minima and maxima.
In the co-moving reference frame of the substrate pattern, $y_\mathrm{com} = y - v_\mathrm{wave} t$, the droplet's velocity $\dot y_\mathrm{com}=\dot y - v_\mathrm{wave}$ generally oscillates around $-v_\mathrm{wave}$, \latin{i.e.},
\begin{equation}
 \dot y_\mathrm{com} = - v_\mathrm{wave} - v_\mathrm{c} \sin\left(\frac{2\pi y_\mathrm{com}}{ \lambda} \right)
 \label{eq_droplet_oscillator}
\end{equation}
with amplitude $v_\mathrm{c} = F_0/\Gamma$.
Depending on its position on the pattern, the droplet's speed is increased or reduced.
In particular, for $y_\mathrm{com} = n \lambda /2$ the force on the droplet is zero, which means $\dot y = 0$ or $\dot y_\mathrm{com} = - v_\mathrm{wave}$.
Although $\dot y_\mathrm{com}$ oscillates around $-v_\mathrm{wave}$, that is not its time average.
Instead $-v_\mathrm{wave}$ is the \emph{spatial} average of $\dot y_\mathrm{com}$.

We can map Eq.~(\ref{eq_droplet_oscillator}) onto the so-called \emph{ideal nonuniform oscillator} for phase variable $\varphi$ with constants $a$ and $c$ \cite{strogatz_nonlinear_1994} by identifying
\begin{equation}
 \varphi = -\frac{2\pi y_\mathrm{com}}{\lambda}\, , \enspace
 a = \frac{2\pi v_\mathrm{wave} }{\lambda}\, , \enspace
 \text{and}~c =\frac{2\pi v_\mathrm{c}}{\lambda} \, .
\end{equation}
This results into the expression
\begin{equation}
 \dot \varphi = a - c \sin \varphi
 \label{eq_nonuniform_oscillator}
\end{equation}
also called the Adler equation~\cite{thiele_driven_2006}.
Unlike the uniform (harmonic) oscillator, the phase of which has a constant rate of change, $\varphi$ determines its own rate of change.
In particular, the oscillation stops for $|c|\geq |a|$ when the two terms on the r.h.s.~balance each other~\cite{adler_study_1946}.

Strogatz derives the period~$T$ of the nonuniform oscillator in Ref.~\cite{strogatz_nonlinear_1994} and following his derivation we find with our parameters
\begin{equation}
 T = \frac{\lambda}{\sqrt{v_\mathrm{wave}^2 - v_\mathrm{c}^2}}~.
 \label{eq_period_nonuniform}
\end{equation}
To predict the long-time behavior of the droplet, for example, its position, the \emph{time} averaged droplet speed $\bar v$ is useful, which we derive now.
Generally, in steady state the droplet position oscillates relative to the substrate pattern with a period
\begin{equation}
 T = \frac{\lambda}{v_\mathrm{wave} - \bar v} \, ,
 \label{eq_period_relative}
\end{equation}
where the difference between $v_\mathrm{wave}$ and  $\bar v$ gives the droplet's speed relative to the substrate pattern with
wavelength~$\lambda$.
Differently speaking, $T$ is the period of the forcing experienced by the droplet through the travelling wave.
Setting Eq.~(\ref{eq_period_nonuniform}) equal to Eq.~(\ref{eq_period_relative}), we obtain
\begin{equation}
 \bar v = v_\mathrm{wave}-\sqrt{v_\mathrm{wave}^2 - v_\mathrm{c}^2} \, ,
\label{eq.mean_v}
\end{equation}
which relates $\bar v$ to $v_\mathrm{wave}$ with only a single free parameter, $v_\mathrm{c}$.
Consequently, in this model $v_\mathrm{c}$ contains all material properties of the driving mechanism and droplet-substrate interaction.
The limiting case for large $v_\mathrm{wave}$ is $\bar v \approx v_\mathrm{c}^2 / (2v_\mathrm{wave})$ and therefore
\begin{equation}
 v_\mathrm{c}=\lim_{v_\mathrm{wave}\to\infty}\sqrt{2v_\mathrm{wave}\bar v} \, ,
 \label{eq_critical_limit}
\end{equation}
which is useful to find $v_\mathrm{c}$ in the simulations but also in experiments.

However, $v_\mathrm{c}$ is not just a fitting parameter because it also marks a change in the dynamics of the droplet.
For $v_\mathrm{wave}\leq v_\mathrm{c}$ the oscillation stops, as predicted by the non-uniform oscillator, which means $\dot y_\mathrm{com}=0$ or, equivalently, $\dot y = v_\mathrm{wave}=\bar v$.
In our following analysis, we will refer to this stationary dynamics as 'surfing' and to the oscillatory dynamics as 'wobbling'.

\section{Travelling waves in wettability}
\label{sec_wettability}
We first study how travelling waves in wettability can move droplets on a completely flat substrate, similar to our earlier study of moving steps in wettability in Ref.~\cite{grawitter_steering_2021}.
However, there are clear differences, which we explain in the following.

\subsection{Substrate dynamic}
Here, we impose a travelling wave in wettability on a flat substrate such that the prescribed equilibrium contact angle varies according to
\begin{equation}
\theta_{\mathrm{eq}}(y, t) = \theta_{0} +
\varTheta(t)\varDelta\theta\sin^2 \left(2\pi\frac{y - v_\mathrm{wave}t}{2\lambda}\right)
\end{equation}
with wavelength $\lambda=2\,R_0$, $v_\mathrm{wave}$ the speed of the wave in $y$-direction, the Heaviside or $\varTheta$-function, the amplitude~$\varDelta\theta=30^\circ$, and $\theta_0=90^\circ$.%
\footnote{%
Note that in Ref.~\cite{grawitter_steering_2021} we studied a moving wettability step and explored the influence of different contact angles $\theta_0$ and step widths $\Delta \theta$ on the surfing speed of a droplet.
One important finding was that $\theta_0=90^\circ$ does not result in a qualitatively different result than other values of $\theta_0$.
}

The droplet moves toward regions of high wettability, which means the valleys of the travelling wave in $\theta_\mathrm{eq}$.
Because these valleys are steadily moving forward in $y$-direction, the droplet is biased to also move in that direction.
However, as noted in Section~\ref{sec_oscillator}, the droplet can perform different kinds of motion under the influence of this driving mechanism, namely surfing and wobbling motion (see Fig.~\ref{fig.wettability_waves}), which we will analyse in detail now.

\subsection{Surfing and Wobbling}
We first review the basic phenomenology.
A surfing droplet moves at the same speed as the travelling wave.
Consequently, its shape remains constant and a stationary flow field exists inside the droplet (see Movie M1 in the ESI).
This stationary state is identical to the surfing dynamics we observed in Ref.~\cite{grawitter_steering_2021}.

A wobbling droplet moves at a slower speed than the travelling wave pattern.
It oscillates in speed and shape as it repeatedly passes over the peaks and valleys of the wave (see Movie M2 in the ESI).
The wobbling dynamics was not present in our previous study of moving steps in wettability~\cite{grawitter_steering_2021} because
it requires a periodic wettability pattern.

According to the nonuniform oscillator model, we outlined in Section~\ref{sec_oscillator}, wobbling occurs above a critical speed of the travelling wave pattern.
As the speed of the travelling wave pattern decreases, the period at which the droplet shape oscillates grows and it diverges exactly at the transition.
So, the resulting surfing state can be considered as a wobbling which is frozen in time.
This bifurcation is analyzed in detail in Section~\ref{sec_sniper}.

\subsection{Quantitative analysis}

\begin{figure}[t]
 \includegraphics[width=\linewidth]{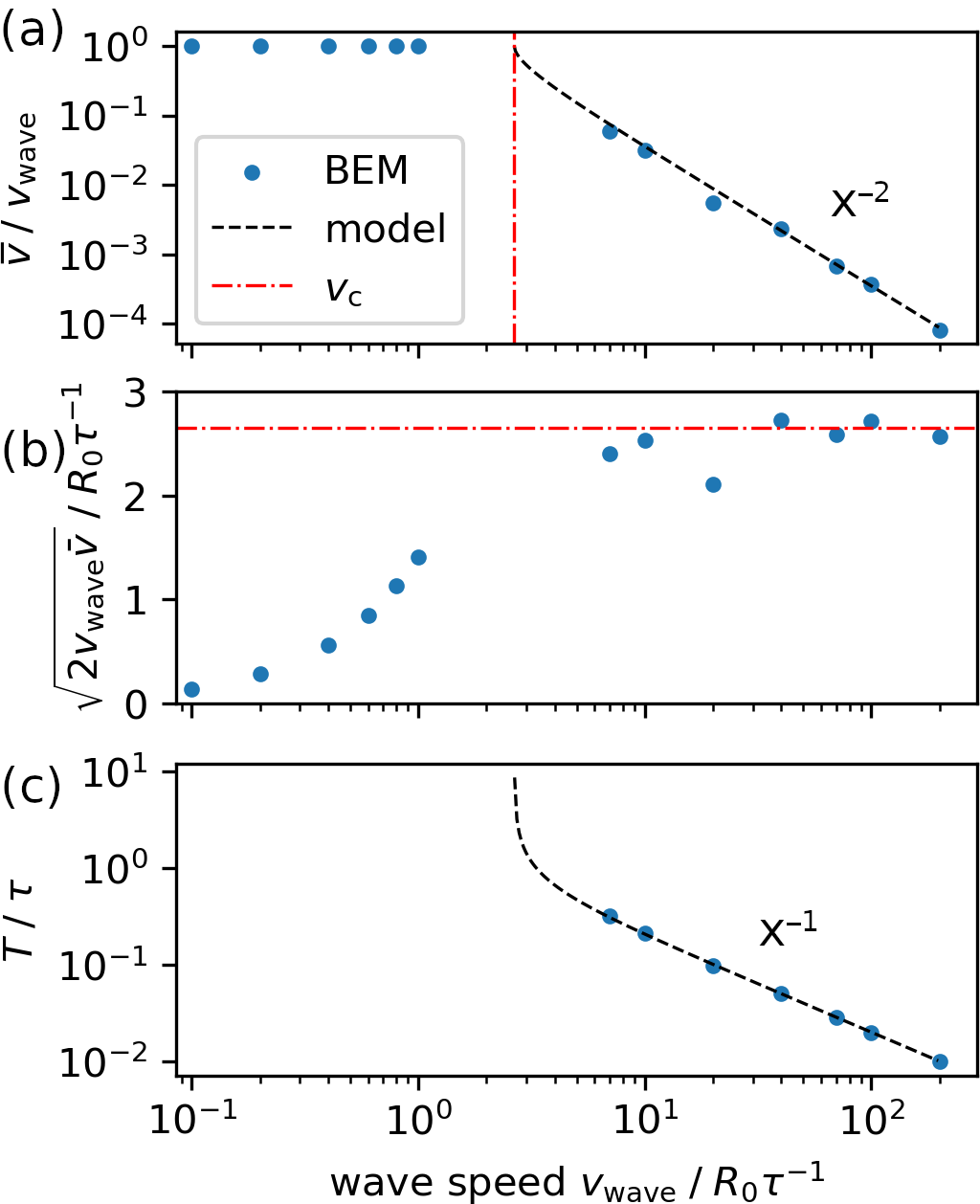}
 \caption{%
 Diagrams of (a)~the mean droplet speed~$\bar v$,
 (b)~the helper quantity $\sqrt{2v_\mathrm{wave}\bar v}$,
 and (c) the wobbling oscillation period~$T$
 plotted versus $v_\mathrm{wave}$ for a travelling wave in wettability with wavelength~$\lambda=2\,R_0$.
 Blue dots indicate our results from BEM simulation, the dashed black lines indicate predictions from the nonuniform oscillator model and the dash-dotted red line indicates the critical wave speed $v_\mathrm{c}$, which is the sole fitting parameter of the model.
}
 \label{fig_wettability}
\end{figure}

One characteristic quantity of the surfing and wobbling dynamics is the droplet speed and we ask how it relates to the speed of the wave pattern.
The droplet speed needs to be defined carefully, especially for wobbling, because it varies periodically over time.
It is appropriate to average the droplet speed over one period of oscillation~$T$.
The period of oscillation refers to the shortest duration between two identical shapes of the droplet, albeit not identical positions in the lab reference frame.
The average droplet speed is then formally given by
\begin{equation}
\bar v =
\frac{y(t + nT) - y(t)}{nT}
\label{eq_avg_speed}
\end{equation}
where the starting time $t$ is set after the decay of any transients and the number of periods $n$ is chosen according to the
available simulation data.
Note that, for the case of a surfing droplet, we have $y(t)=v_\mathrm{wave}t + y_0$ with position $y_0$ at $t=0$.
Here, the time intervall $T$ is arbitrary and the definition above reduces to $\bar v=v_\mathrm{wave}$.

In Fig.~\ref{fig_wettability}(a) we plot the speed ratio $\bar v/v_\mathrm{wave}$ versus $v_\mathrm{wave}$ as determined from our BEM simlations.
The speed ratio is $1$ in the surfing state and it decays as $v_\mathrm{wave}^{-2}$ in the wobbling state (dashed line), as predicted from the nonlinear oscillator model and Eq.~(\ref{eq.mean_v}) in the limit $v_\mathrm{wave} \gg v_c$.
This means $\bar v$ is inversely proportional to $v_\mathrm{wave}$ for wobbling and equal to $v_\mathrm{wave}$ for surfing.
The time-averaged speed of the droplet decays to zero which also matches our observation from an earlier article that droplets are less susceptible to rapid changes in wettability~\cite{grawitter_droplets_2021}.
For a full quantitative comparison with our model, we need to determine the velocity $v_c$.
Following Eq.~(\ref{eq_critical_limit}), we plot $\sqrt{2v_\mathrm{wave}\bar v}$ versus $v_\mathrm{wave}$ in Fig.~\ref{fig_wettability}(b) and read off $v_c = 2.65\,R_0\tau^{-1}$ at large wave speeds (dash-dotted red line).
The dashed line in Fig.~\ref{fig_wettability}(a) is the prediction from Eq.~(\ref{eq.mean_v}) using the determined value for $v_c$.

The transition from surfing to wobbling appears continuous, meaning there is no jump in droplet speed, as predicted by the non-linear oscillator model.
However, the transient dynamics decays slowly close to the transition so that we lack data close to $v_\mathrm{wave}=v_\mathrm{c}$.
If the transition is indeed continuous, the droplet moves fastest exactly at the critical wave speed~$v_\mathrm{c}$.
The period~$T$ of wobbling oscillations, displayed in Figure~\ref{fig_wettability}(c), grows inversely proportional to
$v_\mathrm{wave} -v_\mathrm{c}$ as $v_\mathrm{wave}$ approaches $v_\mathrm{c}$ from above.
These observations for the wobbling state again quantitatively match the prediction of the nonlinear oscillator model.

\subsection{Comparison with travelling steps in wettability}
\label{sec_sniper}
\begin{figure}
 \includegraphics[width=\linewidth]{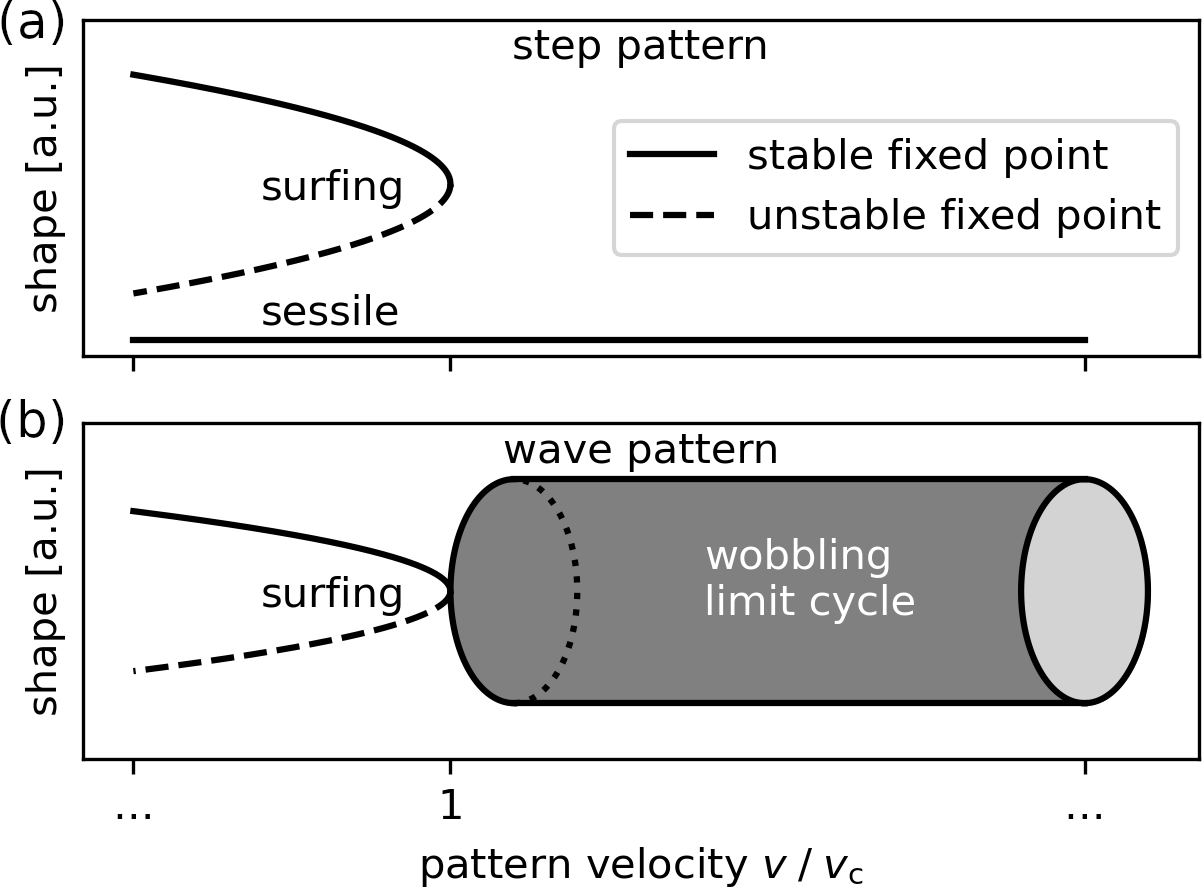}
 \caption{Schematic bifurcation diagrams for an appropriate droplet-shape variable plotted versus pattern velocity.
 (a)~The droplet dynamics in the step wettability pattern (\emph{c.f.} Ref.~\cite{grawitter_steering_2021}) corresponds to a saddle-node bifurcation at the critical velocity $v_\mathrm{c}$.
 (b)~The travelling wave pattern generates an additional limit cycle that collides with the saddle and node, giving rise to a saddle-node-infinite-period (SNIPER) bifurcation.%
}
\label{fig_bifurcations}
\end{figure}

In Ref.~\cite{grawitter_steering_2021} we observed a droplet surfing along a substrate on a wettability step that moves with velocity $v_\mathrm{step}$ and a sigmoidal profile~$\sigma(y)$.
Similar to Eq.~(\ref{eq_droplet_oscillator}) we can describe the droplet velocity in the co-moving frame of the step with
 \begin{equation}
 \dot y_\mathrm{com} = - v_\mathrm{step} + v_\mathrm{c} \sigma'\left(\frac{y_\mathrm{com}}{\delta}\right) \, ,
\end{equation}
where for the sigmoidal function we chose $\sigma(y)=(1 + \mathrm{e}^{-y})^{-1}$, and $\delta$ is the step width.
Most importantly, for $v_\mathrm{step} < v_\mathrm{c}$ two surfing states exist.
They sit outside the center of the step and are stable and unstable fixed points, which eventually meet in a saddle-node bifurcation at $v_\mathrm{step} = v_\mathrm{c}$.
Furthermore, there is a sessile steady state where the droplet sits outside of the step.
It occurs for any value of $v_\mathrm{step}$ in the limit $|y_\mathrm{com}/\delta| \gg 0$ so that the contribution from $\sigma'$ vanishes.
The saddle node bifurcation with the stable and unstable fixed points as well as the stable fixed point of the sessile droplet are schematically illustrated in Fig.~\ref{fig_bifurcations}(a) using an appropriate shape variable.

Comparing the moving wettability step treated in Ref.~\cite{grawitter_steering_2021} to the travelling wettability wave considered here and modeled by Eq.~(\ref{eq_droplet_oscillator}), we note that a sessile state is impossible due to the periodicity of the wave pattern.
Instead, the wobbling dynamic takes the place of the sessile state for $v_\mathrm{wave} > v_\mathrm{c}$.
In Fig.~\ref{fig_bifurcations}(b) the resulting bifurcation diagram is sketched.
The saddle-node bifurcation of the moving wettability step is still present but for travelling waves it collides with the limit cycle associated with wobbling.
This collision comprises a global bifurcation which is called Saddle-Node-Infinite-PERiod (SNIPER) bifurcation~\cite{strogatz_nonlinear_1994}.

A SNIPER bifurcation was previously observed for droplets exposed to an oscillating external force on a \emph{static} heterogeneous substrate~\cite{thiele_driven_2006}
and for a droplet subject to a \emph{constant} external force on a rotating cylinder~\cite{thiele_depinning_2011,thiele_bifurcation_2016}
with the force acting perpendicular to the cylinder's axis of rotation.
In both
cases, the bifurcation marks the transition from a pinned to a sliding droplet.

\section{Travelling-wave deformations}
\label{sec_deformations}

We now turn to the second driving mechanism introduced in Section~\ref{sec_theory}.
The substrate deforms such that its height profile follows a travelling wave.
The wettability is kept uniform and we solely rely on the travelling height profile to move the droplet.
The influence of gravity is negligible because the capillary length $l_\mathrm{c}=\sqrt{\gamma_\mathrm{lg}\rho^{-1}g^{-1}}\approx 23\,R_0$ is much larger than the droplet radius.
This raises the question of the driving mechanism that moves the droplet forward.

For an equilibrium contact angle of $\theta_\mathrm{eq}=90^\circ$, which according to Young's law occurs at surface tensions $\gamma_\mathrm{sl} = \gamma_\mathrm{sg}$, only the liquid-gas interface is relevant and the droplet deforms to minimize the area of this interface.
Hence, on a curved surface the droplet moves away from peaks and toward valleys because there the liquid-gas interface occupies less area compared to a droplet with the same volume sitting on a flat part or a peak of the substrate.
For $\theta_\mathrm{eq} < 90^\circ$ the droplet is even more attracted to valleys in the substrate, as it is energetically advantageous to cover more of the solid with liquid, while for $\theta_\mathrm{eq}  > 90^\circ$ the situation is less clear.
In Ref.~\cite{lv_substrate_2014} Lv~\emph{et al.} demonstrate how a droplet moves on the inside and outside of a conical substrate without any additional external stimulus.
On the outside of the cone the liquid-substrate curvature is negative and on the inside of the cone it is positive.
Notably, for all cases studied with $\theta_\mathrm{eq}$ ranging from $30^\circ$ to $120^\circ$, the droplets move in the direction of increasing curvature.
The same behavior was recently confirmed for droplets on cylinders of varying size~\cite{liu_spontaneous_2022}.
Indeed, when we place a droplet onto a substrate peak with $\theta_\mathrm{eq}=120^\circ$ in our BEM simulations and then perturb the droplet, the position is unstable.
Lv~\emph{et al.} term this mechanism \emph{curvi-propulsion} and in this section we investigate if it generates the same wobbling and surfing dynamics we found for the wettability waves in Section~\ref{sec_wettability} and if there are any differences between the two cases.

\subsection{Surfing and wobbling}
Instead of a wettability profile, we now impose a travelling height profile of the substrate with amplitude~$h_0$,
\begin{equation}
 h(y,t) = h_0q(t)
 \sin^2 \frac{y-\xi - y_\mathrm{wave}(t)}{2 \lambda} \, ,
\end{equation}
where $\xi$ determines the initial phase shift or position of the wave relative to the droplet at $y=0$.
Especially, in the surfing dynamic we use a value of $\xi=0.3\,R_0$ to place the droplet close to its surfing position on the wave profile, while $\xi$ is irrelevant for the wobbling dynamics.
We introduce the factor
\begin{equation}
 q(t) = \begin{cases}
            0&\text{for}~t\leq 0\\
            t/t_0
            & \text{for}~0<t\leq t_0\\
            1 & \text{for}~t>t_0
           \end{cases}
\end{equation}
and
\begin{equation}
 y_\mathrm{wave}(t) = \begin{cases}
      0&\text{for}~t\leq t_0\\
 v_\mathrm{wave}\cdot(t - t_0)&\text{for} t>t_0 \,.
         \end{cases}
\end{equation}
The purpose of the factor $q(t)$ is to allow a continuous evolution of the surface deformation from a flat to the sinusoidal shape to which the droplet can adjust gradually starting at a specific position relative to the wave.
Then, the deformation wave starts to travel at $t=t_0$, where we choose $t_0$ in the interval~$0.1\,\tau\leq t_0\leq0.2\,\tau$.
To be specific, we set $\lambda=4\,R_0$ for the wavelength, $h_0=0.1\,R_0$ for the amplitude, and an equilibrium contact angle of $\theta_\mathrm{eq}=90^\circ$.%
\footnote{Note that $\theta_\mathrm{eq}=90^\circ$ is not a special case here because \emph{curvi-propulsion} acts in the same direction regardless of wettability, \latin{i.e.}, regardless whether a substrate is hydrophobic or hydrophilic~\cite{lv_substrate_2014}.}
In Section~\ref{sec_tuning}, we will explore travelling deformation waves for a range of wavelengths.

With time the droplet again settles into either \emph{surfing} or \emph{wobbling} motion depending on the wave speed $v_\mathrm{wave}$.
Because it is attracted to the valleys in the substrate, it tries to follow them as the travelling deformation wave progresses.
If the travelling wave moves sufficiently slowly, the droplet can surf on a position behind the valley (see Movie M3 in the ESI).
If the travelling wave moves faster than a critical speed, the droplet is always overtaken by the peak behind it, relaxes into the closest valley, and falls back again giving rise to the wobbling motion (see Movie M4 in the ESI).

\subsection{Quantitative analysis}

\begin{figure}
 \includegraphics[width=\linewidth]{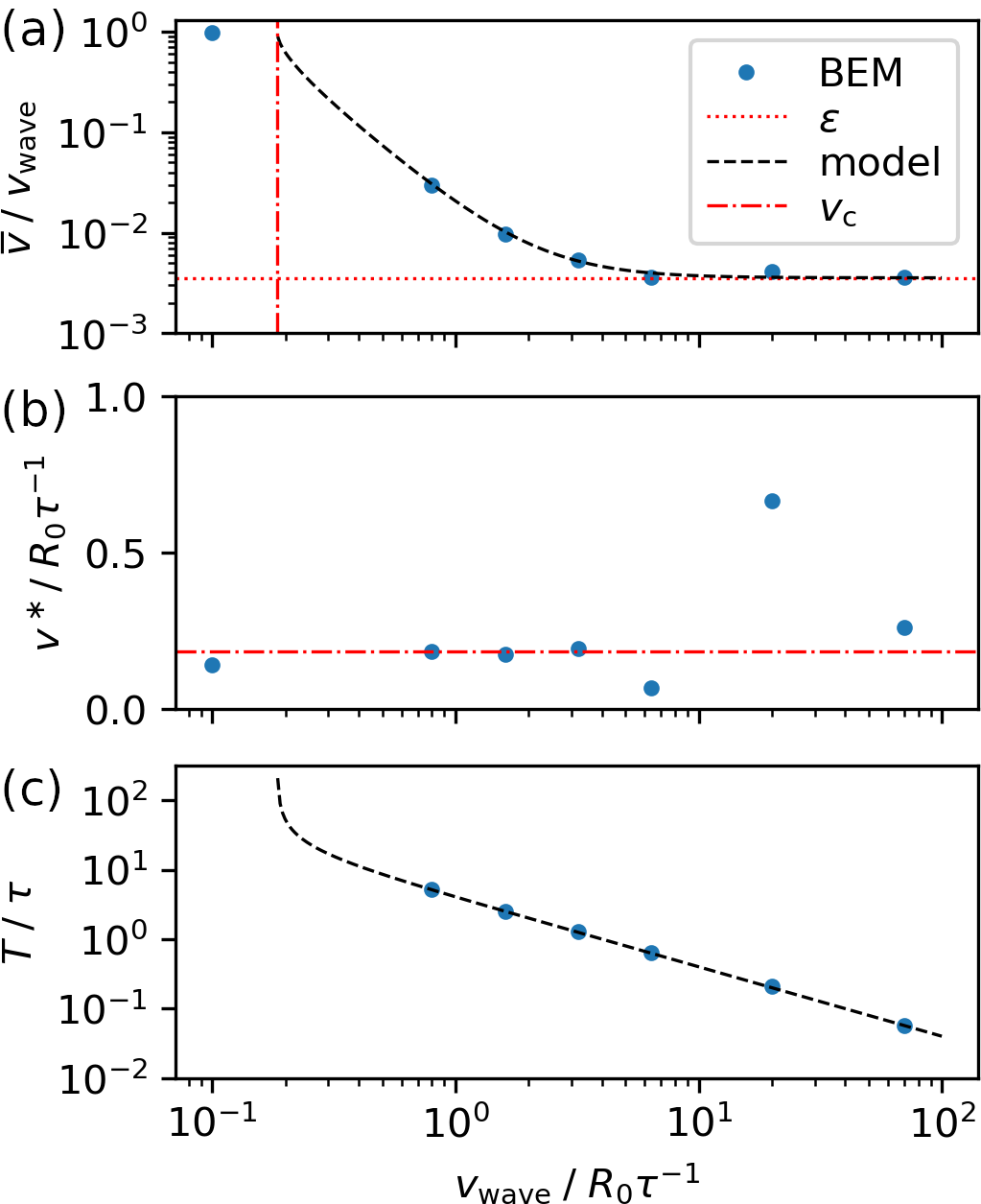}
  \caption{%
 Diagrams of (a) the mean droplet speed~$\bar v$, (b)  the helper quantity $v^\ast$, and (c) the wobbling oscillation period~$T$ as functions of $v_\mathrm{wave}$ for a travelling deformation wave with wavelength~$\lambda=4\,R_0$.
 Blue dots indicate our results from BEM simulation, the dashed black line indicates the revised nonuniform oscillator model, Eq.~(\ref{eq_revised_model}), the dash-dotted red line indicates the critical wave speed $v_\mathrm{c}$, and the dotted red line indicates the parameter $\varepsilon$.
 The latter two are the fitting parameters of the model.%
}
 \label{fig_deformation_combined}
\end{figure}

In analogy to the travelling wettability pattern, we analyze the time-averaged speed of the droplet as defined in Eq.~(\ref{eq_avg_speed}).
In Fig.~\ref{fig_deformation_combined}(a) we again plot $\bar{v} / v_\mathrm{wave}$ versus $v_\mathrm{wave}$.
In contrast to the wettability wave, the speed of the wobbling droplet now grows linearly with $v_\mathrm{wave}$ at large $v_\mathrm{wave}$.
Due to this clear difference, we realize that our model needs to be adjusted.

To start, we re-examine the nonlinear oscillator model of Eq.~(\ref{eq_droplet_oscillator}), where a spatially periodic force drives the droplet with a constant amplitude $v_\mathrm{c}$.
For the deforming substrate we introduced a traction force $\vec F_\bot$ in Eq.~(\ref{eq_force_balance_complete}), which imposes the deformation of the substrate onto the droplet shape.
However, this force depends on $v_\mathrm{wave}$:
It is zero for $v_\mathrm{wave}\leq v_\mathrm{c}$, where the droplet's shape is constant and beyond $v_\mathrm{c}$ it grows with $v_\mathrm{wave}$ as the droplet is deformed more rapidly.
As before, the force is also spatially periodic since it originates from the wave pattern of the substrate's height profile.
Therefore, to include the dependence of the amplitude of the driving force on $v_\mathrm{wave}$, we extend Eq.~(\ref{eq_droplet_oscillator}) and write
\begin{equation}
  \dot y_\mathrm{com} = - v_\mathrm{wave} -
  v_c
  \left[ 1 +
  \varepsilon\frac{v_\mathrm{wave}^{2}-v_\mathrm{c}^{2}}{ v_\mathrm{c}^2  }\right]\sin\left(\frac{2\pi y_\mathrm{com}}{\lambda}\right) \, .
  \label{eq_droplet_oscillator_revised}
\end{equation}
The dimensionless coefficient $\varepsilon\ll 1$ quantifies the relative strength of the wave-speed depending part of the amplitude.
In analogy to Eq.~(\ref{eq_period_nonuniform}), we find for the time period,
\begin{equation}
 T=\frac{\lambda}{\sqrt{1-2\varepsilon
 -\varepsilon^2(v_\mathrm{wave}^2-v_\mathrm{c}^2)/v_\mathrm{c}^2
 }}\cdot\frac{1}{\sqrt{v_\mathrm{wave}^2-v_\mathrm{c}^2
 }} \,,
 \label{eq_period_nonuniform_revised}
\end{equation}
which together with Eq.~(\ref{eq_period_relative}) gives the mean droplet speed,
\begin{equation}
 \bar v = v_\mathrm{wave}-(1-\varepsilon)\sqrt{v_\mathrm{wave}^2 - v_\mathrm{c}^2} \, .
 \label{eq_revised_model}
\end{equation}
Here, we have used $\sqrt{1-2\varepsilon-\varepsilon^2\ldots}\approx1-\varepsilon$ in Eq.~(\ref{eq_period_nonuniform_revised}).
The small parameter $\varepsilon$ means that $\bar v$ increases proportional to $v_\mathrm{wave}$ for large $v_\mathrm{wave}$.
Note that $\varepsilon$ does not affect the scaling of the wobbling period~$T$ with the inverse wave speed well above $v_c$.

Now, our procedure to determine $\varepsilon$ and $v_\mathrm{c}$, the central parameters of the model, is as follows.
First, from Eq.~(\ref{eq_revised_model}) we determine the limiting value of $\bar v / v_\mathrm{wave} = \varepsilon$ for large $v_\mathrm{wave}$, which we can directly read off from Fig.~\ref{fig_deformation_combined}(a) as $\varepsilon=3.565\cdot 10^{-3}$.
Then, we subtract $\varepsilon v_\mathrm{wave}$ from both sides of Eq.~(\ref{eq_revised_model}),
\begin{equation}
 \bar v - \varepsilon v_\mathrm{wave}=(1-\varepsilon)[v_\mathrm{wave}-\sqrt{v_\mathrm{wave}^2-v_\mathrm{c}^2}] \, ,
\end{equation}
and expand the r.h.s.~for large $v_\mathrm{wave}$ to obtain
\begin{equation}
  \bar v - \varepsilon v_\mathrm{wave}=\lim\limits_{v_\mathrm{wave}\gg v_\mathrm{c}}(1-\varepsilon)\frac{v_\mathrm{c}^2}{2v_\mathrm{wave}}~.
\end{equation}
Solving for $v_\mathrm{c}$ yields
\begin{equation}
 v_\mathrm{c}=\lim\limits_{v_\mathrm{wave}\gg v_\mathrm{c}} \sqrt{2v_\mathrm{wave} \frac{\bar v - \varepsilon v_\mathrm{wave}}{1 - \varepsilon}}
 = \lim\limits_{v_\mathrm{wave}\gg v_\mathrm{c}} v^\ast  \, ,
\label{eq.vc_limit}
\end{equation}
which for $\varepsilon=0$ indeed simplifies to Eq.~(\ref{eq_critical_limit}).
Accordingly, in Fig.~\ref{fig_deformation_combined}(b) we plot $v^\ast$ as defined in Eq.~(\ref{eq.vc_limit}).
Because $\varepsilon$ appears as coefficient of $v_\mathrm{wave}^2$, $v^\ast$ is sensitive to small errors in $\varepsilon$ at large $v_\mathrm{wave}$.
However, around $v_\mathrm{wave} / R_0\tau^{-1} = 2$ it takes on a constant value, from which we determine $v_\mathrm{c}=0.185\,R_0\tau^{-1}$.
The dashed black line labelled \emph{model} in Fig.~\ref{fig_deformation_combined}(a) shows that with these parameters our revised model quantitatively matches the BEM data (blue dots).
Similarly, it also reproduces the wobbling period displayed in Fig.~\ref{fig_deformation_combined}(c).
Because around $v_\mathrm{wave} = v_\mathrm{c}$, the original and the revised model behave the same, the latter also predicts a SNIPER bifurcation at $v_\mathrm{wave} = v_\mathrm{c}$ for droplets driven by travelling substrate deformations.

Our revised model fits the BEM data well and thus provides an interpretation of the asymptotic behavior of $\bar v$ for large $v_\mathrm{wave}$.
To capture the asymptote, we introduced an amplitude of the driving force, which depends on $v_\mathrm{wave}$.
It accounts for the fact that a travelling substrate deformation always needs to displace the droplet, which requires a larger driving force with inreasing $v_\mathrm{wave}$.
This is the clear difference to travelling wettability patterns treated in
Section~\ref{sec_wettability}, where the pattern can just pass beneath the droplet without the need to lift it up.

\section{A note on particle sorting}
\label{sec_tuning}
\begin{figure}
 \includegraphics[width=\linewidth]{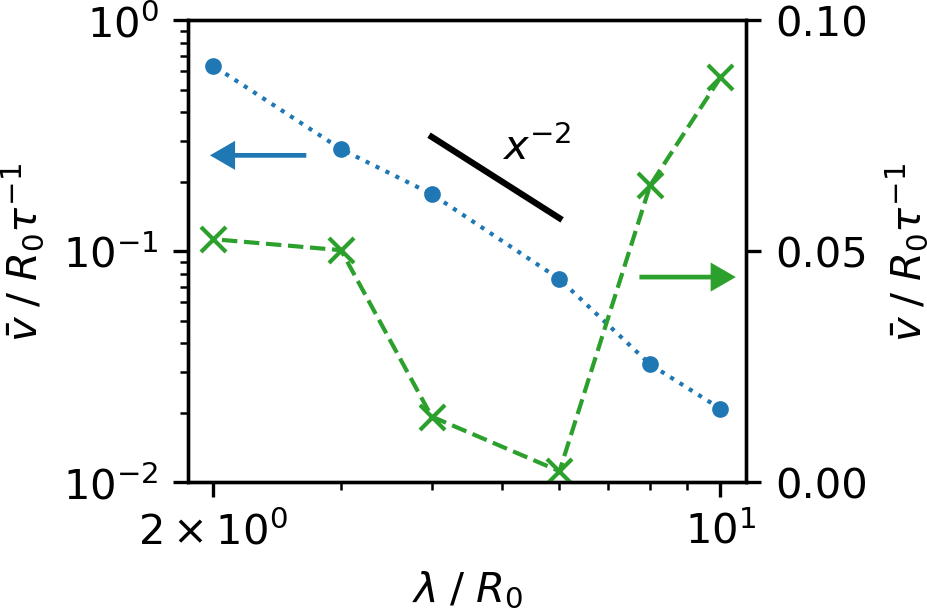}
 \caption{%
Diagram of mean droplet speed~$\bar v$ as a function of $\lambda$ for travelling waves with speed~$v_\mathrm{wave}=50\,R_0\tau^{-1}$.
Blue dots indicate results from BEM simulation with waves in wettability (left axis), and green crosses indicate results from BEM simulation with deformation waves (right axis).
Connecting dotted and dashed lines are added to guide the eye.
The solid black line indicates an inverse-square-law scaling.
}
 \label{fig_wavelength_speed}
\end{figure}
In Sections~\ref{sec_wettability} and \ref{sec_deformations} we described that droplets move in response to spatial gradients in wettability or curvature of the substrate, respectively.
Both, gradient and curvature, change in magnitude when we adjust the wavelength~$\lambda$ of the travelling wave pattern.
Therefore, we can expect that the speed of the droplet~$\bar v$ also depends on $\lambda$.

In Fig.~\ref{fig_wavelength_speed} we display for both driving mechanisms the mean droplet velocity $\bar v$ as a function of $\lambda$ for $v_\mathrm{wave}=50\,R_0\tau^{-1}$, which places the droplet far into the wobbling dynamic.
For travelling deformation waves (green crosses in Fig.~\ref{fig_wavelength_speed}) no clear trend is discernible.
However, for travelling waves in wettability (blue dots in Fig.~\ref{fig_wavelength_speed}) we observe $\bar v$ decays as $(\lambda/R_0)^{-2}$.
Recall from Eq.~(\ref{eq_critical_limit}) that the asymptotic behavior at large $v_\mathrm{wave}$ is $\bar v=v_\mathrm{c}^2/2v_\mathrm{wave}$.
Since $v_\mathrm{wave}$ is independent of $\lambda$, we can infer that $v_\mathrm{c}\sim(\lambda/R_0)^{-1}$.
Consequentially, the ratio $\lambda/R_0$ can be used to tune at which speed~$v_\mathrm{c}$ the transition from surfing to wobbling occurs.

Now consider a poly-disperse collection of droplets, such that $R_0$ varies among them.
This means, for each droplet size there is a specific $v_\mathrm{c}$ and some droplets might be in the surfing regime and others in the wobbling regime at the same value of $v_\mathrm{wave}$.
Since droplets generally move more slowly in the wobbling regime than in the surfing regime, this naturally leads to a sorting mechanism for poly-disperse collections of droplets.

\section{Conclusions}
\label{sec_conclusions}

We have studied liquid-droplet motion on substrates using two different mechanisms: travelling waves in wettability and travelling-wave deformations of the substrate.
To that end, we first implemented the boundary element method (BEM) for the fluid velocity field inside the droplet starting from the free energy of the droplet,
the balance of forces at the droplet interface, and the relevant friction contributions.

To interpret our numerical findings from applying the BEM, we first introduced an analytic model for a particle driven by a travelling-wave
force, which maps onto the so-called nonuniform oscillator. For increasing wave velocity, the analytic model exhibits a bifurcation between
two states, which we call \emph{surfing} and \emph{wobbling} and which directly connect to the phenomenology observed in our BEM
simulations.

For travelling waves in wettability the results from BEM simulations for wobbling period and mean droplet speed are consistent with the analytic model and indeed confirm the bifurcation in the full dynamical system.
This further implies that the motion of the droplet's center of mass can be characterized
solely by the critical wave speed for the bifurcation, which is determined by the material properties of the droplet and substrate.
For a specific set of parameters we determined the critical wave speed by varying the wave speed in the BEM simulations.

We also investigated travelling-wave deformations of the substrate, which give rise to what is called curvi-propulsion in
Ref.~\cite{lv_substrate_2014}.
Again, we could identify a transition from surfing to wobbling with increasing wave speed but, interestingly, the simulation results reveal that the droplet speed does not approach zero for large wave speeds.
By including a speed-dependent amplitude of the driving force, the modified analytic model could account for this new behavior, as we demonstrated for a specific set of parameters in the BEM simulations.

Finally, we also performed simulations where we varied the wavelength $\lambda$ of the travelling wave for both driving mechanisms in the wobbling regime.
For waves in wettability the droplet speed decays as $\lambda^{-2}$, which implies that the critical wave speed is proportional to droplet radius over wavelength.
This suggests a potential sorting mechanism for a collection of droplets by droplet size.
For traveling-wave deformations of the substrate, we could not observe a clear trend in droplet speed w.r.t.~wavelength.

Our findings compare and contrast two different mechanisms for steering droplets toward a preferred direction, one based on switchable
wettability and one based on externally-induced deformations of the substrate. They demonstrate that the relevant quantities such as
droplet speed and wobbling period can be reproduced and interpreted with an analytic model using only one or two fitting parameters,
respectively.
Moving droplets is relevant for lab-on-a-chip applications. Our work suggest two methods for realizing this.
Another attractive possibility to explore are substrates with switchable softness \cite{nekoonam_controllable_2023} and the occurence of durotaxis.

\section*{Conflicts of interest}
There are no conflicts to declare.

\section*{Acknowledgements}
We thank Uwe Thiele and Akash Choudhary for fruitful discussions and acknowledge financial support from German Research Foundation (DFG) as part of priority program~2171 (project number 505839720).

\appendix
\section{Invariance of the boundary-integral equation}
\label{app_pressure}
In Eq.~(\ref{eq_boundary_integrals}) one can replace $\vec {\sigma n}$ by  $\vec{\tilde \sigma n} = \vec {\sigma n} + p_0 \vec n$, where the constant pressure $p_0$ does not contribute to the surface integral.
To show this, we write for an arbitrary force vector $\vec f_0$,
\begin{multline}
\vec f_0 \cdot \oint \vec O(
 \vec{r}_0-\vec{r}
 )
  p_0\vec{  n} \dif^2\vec r
  =\\ p_0  \oint [\vec O(
 \vec{r}-\vec{r}_0
 ) \vec f_0 ] \cdot
  \vec{n} \dif^2\vec r~.
\end{multline}
The closed surface integral of the Stokeslet flow field, $\vec u_\mathrm{Stokeslet}(\vec {r})=\vec O( \vec{r}-\vec{r}_0)\vec f_0$, vanishes, since $\vec u_\mathrm{Stokeslet}(\vec {r})$ is incompressible by construction and after applying Gauss's theorem.

\section{Normal vector in Monge representation}
\label{app_normal}
We parameterize the substrate according to Monge representation by $\vec {\tilde r}=(x, y, h(x, y))$.
The normal~$\vec n$ then points perpendicular to both tangential vectors of the surface which are given by the derivatives of $\vec{\tilde r}$ w.r.t.~$x$ and $y$, respectively, so that
\begin{equation}
 \vec n = \frac{\partial \vec{\tilde r}}{\partial x} \times \frac{\partial \vec{\tilde r}}{\partial y}=
 \begin{pmatrix}
  -\partial_x h\\
  -\partial_y h\\
  1
 \end{pmatrix}~.
\end{equation}
Note that $\vec n$, as written here, is generally not of unit length.
However, as noted in Section~\ref{sec_lagrange}, the normalization constant is subsumed into the Lagrange multiplier~$\lambda$.

\section{Block matrices}
\label{app_block_matrices}
The matrices in Eq.~(\ref{eq.bem}) are defined as
\begin{equation}
 X_{ij}=\int_{C_j} \vec O(\vec r_i - \vec r) \,\mathrm{d}^2\vec r
\end{equation}
and
\begin{equation}
 Y_{ij}=\int_{C_j} \vec T(\vec r_i - \vec r) \vec n \,\mathrm{d}^2\vec r \, ,
\end{equation}
where the surface integral is performed over the polygonal cell $C_j$ around vertex $j$ and $\vec r_i$ is the position of vertex $i$.
Furthermore, in Eq.~(\ref{eq.bem}), if vertex $i$ is on the contact line, $c_i=\theta_i / 2\pi$ with contact angle $\theta_i$ and otherwise $c_i=1/2$.

\section{Supplementary Material}
\label{app_supplementary}
Supplementary material consists of four animations, M1--M4, corresponding to specific BEM simuluations:
M1 and M2 correspond to simulations with $v_\mathrm{wave}=1\,R_0\tau^{-1}$ (surfing) and $v_\mathrm{wave}=10\,R_0\tau^{-1}$
(wobbling) in Fig.~\ref{fig_wettability}, respectively. M3 corresponds to a simulation as described in Section~\ref{sec_deformations} with
$\xi=1\,R_0$, $t_0=0.2\,\tau$ and $v_\mathrm{wave}=0.18\,R_0\tau^{-1}$ (surfing).
M4 corresponds to the simulation with $v_\mathrm{wave}=70\,R_0\tau^{-1}$ (wobbling) in Fig.~\ref{fig_deformation_combined}.

\bibliography{references}

\begin{thebibliography}{44}

\bibitem{alwazzan_condensation_2017}
M.\!\, Alwazzan, K.\!\, Egab, B.\!\, Peng, J.\!\, Khan and C.\!\, Li, Int. J.
  Heat Mass Transf. \textbf{112}, 991 (2017).

\bibitem{edalatpour_managing_2018}
M.\!\, Edalatpour, L.\!\, Liu, A.\!\, Jacobi, K.\!\, Eid and A.\!\, Sommers,
  Appl. Energy \textbf{222}, 967 (2018).

\bibitem{varanakkottu_light_2016}
S.\,N.\!\, Varanakkottu, M.\!\, Anyfantakis, M.\!\, Morel, S.\!\, Rudiuk and
  D.\!\, Baigl, Nano Lett. \textbf{16}, 644 (2016).

\bibitem{qi_mechanical_2019}
L.\!\, Qi, Y.\!\, Niu, C.\!\, Ruck and Y.\!\, Zhao, Lab Chip \textbf{19}, 223
  (2019).

\bibitem{bonn_wetting_2009}
D.\!\, Bonn, J.\!\, Eggers, J.\!\, Indekeu, J.\!\, Meunier and E.\!\, Rolley,
  Rev. Mod. Phys. \textbf{81}, 739 (2009).

\bibitem{teng_recent_2020}
P.\!\, Teng, D.\!\, Tian, H.\!\, Fu and S.\!\, Wang, Mater. Chem. Front.
  \textbf{4}, 140 (2020).

\bibitem{thiele_sliding_2002}
U.\!\, Thiele, K.\!\, Neuffer, M.\!\, Bestehorn, Y.\!\, Pomeau and M.\,G.\!\,
  Velarde, Colloids Surf. A \textbf{206}, 87 (2002).

\bibitem{baigl_photo_2012}
D.\!\, Baigl, Lab Chip \textbf{12}, 3637 (2012).

\bibitem{venancio_digital_2014}
A.\!\, Venancio-Marques and D.\!\, Baigl, Langmuir \textbf{30}, 4207 (2014).

\bibitem{chaudhury_how_1992}
M.\,K.\!\, Chaudhury and G.\,M.\!\, Whitesides, Science \textbf{256}, 1539
  (1992).

\bibitem{ichimura_light_2000}
K.\!\, Ichimura, S.\!\, Oh and M.\!\, Nakagawa, Science \textbf{288}, 1624
  (2000).

\bibitem{vialetto_magnetic_2017}
J.\!\, Vialetto, M.\!\, Hayakawa, N.\!\, Kavokine, M.\!\, Takinoue, S.\,N.\!\,
  Varanakkottu, S.\!\, Rudiuk, M.\!\, Anyfantakis, M.\!\, Morel and D.\!\,
  Baigl, Angew. Chem. Int. Ed. \textbf{56}, 16565 (2017).

\bibitem{schiphorst_light_2018}
J.\!\, ter Schiphorst, J.\!\, Saez, D.\!\, Diamond, F.\!\, Benito-Lopez and
  A.\,P.\,H.\,J.\!\, Schenning, Lab Chip \textbf{18}, 699 (2018).

\bibitem{grawitter_steering_2021}
J.\!\, Grawitter and H.\!\, Stark, Soft Matter \textbf{17}, 2454 (2021).

\bibitem{eral_contact_2013}
H.\,B.\!\, Eral, D.\,J.\,C.\,M.\!\, {’t Mannetje} and J.\,M.\!\, Oh, Colloid
  Polym. Sci. \textbf{291}, 247 (2013).

\bibitem{lim_photoreversibly_2006}
H.\,S.\!\, Lim, J.\,T.\!\, Han, D.\!\, Kwak, M.\!\, Jin and K.\!\, Cho, J. Am.
  Chem. Soc. \textbf{128}, 14458 (2006).

\bibitem{karpitschka_liquid_2016}
S.\!\, Karpitschka, A.\!\, Pandey, L.\,A.\!\, Lubbers, J.\,H.\!\, Weijs, L.\!\,
  Botto, S.\!\, Das, B.\!\, Andreotti and J.\,H.\!\, Snoeijer, Proceedings of
  the National Academy of Sciences \textbf{113}, 7403 (2016).

\bibitem{lv_substrate_2014}
C.\!\, Lv, C.\!\, Chen, Y.\,C.\!\, Chuang, F.\,G.\!\, Tseng, Y.\!\, Yin, F.\!\,
  Grey and Q.\!\, Zheng, Phys. Rev. Lett. \textbf{113}, 026101 (2014).

\bibitem{palagi_structured_2016}
S.\!\, Palagi, A.\,G.\!\, Mark, S.\,Y.\!\, Reigh, K.\!\, Melde, T.\!\, Qiu,
  H.\!\, Zeng, C.\!\, Parmeggiani, D.\!\, Martella, A.\!\, Sanchez-Castillo,
  N.\!\, Kapernaum et~al., Nat. Mater. \textbf{15}, 647 (2016).

\bibitem{rehor_photoresponsive_2021}
I.\!\, Rehor, C.\!\, Maslen, P.\,G.\!\, Moerman, B.\,G.\!\, van Ravensteijn,
  R.\!\, van Alst, J.\!\, Groenewold, H.\,B.\!\, Eral and W.\,K.\!\, Kegel,
  Soft Robotics \textbf{8}, 10 (2021).

\bibitem{gelfand_wetting_1987}
M.\,P.\!\, Gelfand and R.\!\, Lipowsky, Phys. Rev. B \textbf{36}, 8725 (1987).

\bibitem{duprat_elastocapillarity_2016}
C.\!\, Duprat and H.\,A.\!\, Stone, in \emph{Fluid-Structure Interactions in
  Low-Reynolds-Number Flows} (The Royal Society of Chemistry, 2016), pp.
  193--246, ISBN 978-1-84973-813-2.

\bibitem{bico_elastocapillarity_2018}
J.\!\, Bico, {\' E}.\!\, Reyssat and B.\!\, Roman, Annu. Rev. Fluid Mech.
  \textbf{50}, 629 (2018).

\bibitem{aland_unified_2021}
S.\!\, Aland and D.\!\, Mokbel, Int. J. Numer. Methods Eng. \textbf{122}, 903
  (2021).

\bibitem{shuttleworth_surface_1950}
R.\!\, Shuttleworth, Proc. Phys. Soc. Sect. A \textbf{63}, 444 (1950).

\bibitem{andreotti_statics_2020}
B.\!\, Andreotti and J.\,H.\!\, Snoeijer, Annu. Rev. Fluid Mech. \textbf{52},
  285 (2020).

\bibitem{strogatz_nonlinear_1994}
S.\!\, Strogatz, \emph{Nonlinear dynamics and Chaos} (Westview Press, 1994).

\bibitem{pozrikidis_boundary_1992}
C.\!\, Pozrikidis, \emph{Boundary integral and singularity methods for
  linearized viscous flow} (Cambridge University Press, 1992).

\bibitem{grawitter_droplets_2021}
J.\!\, Grawitter and H.\!\, Stark, Soft Matter \textbf{17}, 9469 (2021).

\bibitem{doi_onsager_2011}
M.\!\, Doi, J. Phys.: Condens. Matter \textbf{23}, 284118 (2011).

\bibitem{zafferi_generic_2023}
A.\!\, Zafferi, D.\!\, Peschka and M.\!\, Thomas, Z. Angew. Math. Mech.
  \textbf{103}, e202100254 (2023).

\bibitem{katsikadelis_gauss_2016}
in \emph{The Boundary Element Method for Engineers and Scientists (Second
  Edition)}, edited by J.\,T.\!\, Katsikadelis (Academic Press, Oxford, 2016),
  pp. 403--421, second edition~edn., ISBN 978-0-12-804493-3.

\bibitem{deckelnick_dziuk_elliott_2005}
K.\!\, Deckelnick, G.\!\, Dziuk and C.\,M.\!\, Elliott, Acta Numerica
  \textbf{14}, 139–232 (2005).

\bibitem{huh_hydrodynamic_1971}
C.\!\, Huh and L.\,E.\!\, Scriven, J. Colloid Interface Sci. \textbf{35}, 85
  (1971).

\bibitem{voinov_hydrodynamics_1976}
O.\,V.\!\, Voinov, Fluid Dyn. \textbf{11}, 714 (1976).

\bibitem{deruijter_contact_1997}
M.\,J.\!\, de~Ruijter, J.\!\, De~Coninck, T.\,D.\!\, Blake, A.\!\, Clarke and
  A.\!\, Rankin, Langmuir \textbf{13}, 7293 (1997).

\bibitem{bolanos_derivation_2017}
S.\,J.\!\, Bolaños and B.\!\, Vernescu, Phys. Fluids \textbf{29}, 057103
  (2017).

\bibitem{moffatt_viscous_1964}
H.\,K.\!\, Moffatt, J. Fluid Mech. \textbf{18}, 1 (1964).

\bibitem{thiele_driven_2006}
U.\!\, Thiele and E.\!\, Knobloch, Phys. Rev. Lett. \textbf{97}, 204501 (2006).

\bibitem{adler_study_1946}
R.\!\, Adler, Proc. IRE \textbf{34}, 351 (1946).

\bibitem{thiele_depinning_2011}
U.\!\, Thiele, J. Fluid Mech. \textbf{671}, 121 (2011).

\bibitem{thiele_bifurcation_2016}
T.\,S.\!\, Lin, S.\!\, Rogers, D.\!\, Tseluiko and U.\!\, Thiele, Phys. Fluids
  \textbf{28}, 082102 (2016).

\bibitem{liu_spontaneous_2022}
J.\!\, Liu, Z.\!\, Feng, W.\!\, Ouyang, L.\!\, Shui and Z.\!\, Liu, ACS Omega
  \textbf{7}, 20975 (2022).

\bibitem{nekoonam_controllable_2023}
N.\!\, Nekoonam, G.\!\, Vera, A.\!\, Goralczyk, F.\!\, Mayoussi, P.\!\, Zhu,
  D.\!\, Böcherer, A.\!\, Shakeel and D.\!\, Helmer, ACS Appl. Mater.
  Interfaces \textbf{15}, 27234 (2023).

\end{thebibliography}

\end{document}